\documentclass[apj]{emulateapj}
\usepackage{comment}

\newcommand{\ctbd}[1]{}

\newcommand{\lc}{light curve}
\newcommand{\lcs}{light curves}
\newcommand{\Lc}{Light curve}

\newcommand{\kms}{\ensuremath{\rm km\,s^{-1}}}
\newcommand{\ms}{\ensuremath{\rm m\,s^{-1}}}

\newcommand{\gcmc}{\ensuremath{\rm g\,cm^{-3}}}

\newcommand{\logg}{\ensuremath{\log{g}}}
\newcommand{\vsini}{\ensuremath{v \sin{i}}}

\newcommand{\rsun}{\ensuremath{R_\sun}}
\newcommand{\msun}{\ensuremath{M_\sun}}

\newcommand{\rstar}{\ensuremath{R_\star}}

\newcommand{\teffstar}{\ensuremath{T_{\rm eff}}}

\newcommand{\rpl}{\ensuremath{R_{p}}}

\newcommand{\rjup}{\ensuremath{R_{\rm J}}}
\newcommand{\mjup}{\ensuremath{M_{\rm J}}}

\newcommand{\rjuplong}{\ensuremath{R_{\rm Jup}}}
\newcommand{\mjuplong}{\ensuremath{M_{\rm Jup}}}

\newcommand{\secr}[1]{\mbox{\S\ \ref{sec:#1}}}

\newcommand{\flwof}{\mbox{FLWO 1.2 m}}

\newcommand{\flwos}{\mbox{FLWO 1.5 m}}

\newcommand{\hj}{hot Jupiter}
\newcommand{\vhj}{very hot Jupiter}

\newcommand{\band}[1]{\ensuremath{#1}-band}

\defcitealias{kovacs05}{KBN05}
\defcitealias{fortney06}{FMB06}

\newcommand{\hatcur}{HAT-P-7}
\newcommand{\hatcurb}{HAT-P-7b}

\newcommand{\hatcurCCra}{\ensuremath{19^{\mathrm{h}}28^{\mathrm{m}}59^{\mathrm{s}}.35}}	%
\newcommand{\hatcurCCdec}{\ensuremath{+47^{\circ}58'10''.2}}		%
\newcommand{\hatcurCCmag}{\ensuremath{9.85}}
\newcommand{\hatcurCCtwomass}{2MASS~19285935+4758102}
\newcommand{\hatcurCCgsc}{GSC~03547-01402}
\newcommand{\hatcurCCtassmv}{10.51}
\newcommand{\hatcurCCtassmvshort}{10.5}
\newcommand{\hatcurCCtassvi}{\ensuremath{0.60\pm0.07}}

\newcommand{\hatcurLCdip}{\ensuremath{7.0}}				
\newcommand{\hatcurLCrprstar}{\ensuremath{0.0763\pm0.0010}}		%
\newcommand{\hatcurLCimp}{\ensuremath{0.37^{+0.15}_{-0.29}}}		%
\newcommand{\hatcurLCdur}{\ensuremath{0.1685\pm0.0110}}			%
\newcommand{\hatcurLCingdur}{\ensuremath{0.0150\pm 0.0036}}		%
\newcommand{\hatcurLCP}{\ensuremath{2.2047299\pm0.0000040}}		%
\newcommand{\hatcurLCPprec}{\ensuremath{2.2047299}}			%
\newcommand{\hatcurLCPshort}{2.2047}					%
\newcommand{\hatcurLCT}{\ensuremath{2,453,790.2593\pm0.0010}}		
\newcommand{\hatcurLCMT}{\ensuremath{53,790.2593\pm0.0010}}		%

\newcommand{\hatcurSMEteff}{\ensuremath{6350\pm80}}			%
\newcommand{\hatcurSMEzfeh}{\ensuremath{+0.26\pm0.08}}			%
\newcommand{\hatcurSMElogg}{\ensuremath{4.06\pm0.10}}			%
\newcommand{\hatcurSMEvsin}{\ensuremath{3.8\pm0.5}}			%

\newcommand{\hatcurYYm}{\ensuremath{1.47^{+0.08}_{-0.05}}}		%
\newcommand{\hatcurYYr}{\ensuremath{1.84^{+0.23}_{-0.11}}}		%
\newcommand{\hatcurYYlogg}{\ensuremath{4.07^{+0.04}_{-0.08}}}		%
\newcommand{\hatcurYYlum}{\ensuremath{4.9^{+1.5}_{-0.6}}}		%
\newcommand{\hatcurYYmv}{\ensuremath{3.00\pm0.22}}			%
\newcommand{\hatcurYYage}{\ensuremath{2.2\pm1.0}}			%
\newcommand{\hatcurYYspec}{F6}

\newcommand{\hatcurRVK}{\ensuremath{213.5\pm1.9}}				%
\newcommand{\hatcurRVgamma}{\ensuremath{-37.0\pm1.5}}

\newcommand{\hatcurPPi}{\ensuremath{85\fdg7^{+3.5}_{-3.1}}}		%
\newcommand{\hatcurPPlogg}{\ensuremath{3.31\pm0.08}}			%
\newcommand{\hatcurPPar}{\ensuremath{4.35^{+0.28}_{-0.38}}}			%
\newcommand{\hatcurPParel}{\ensuremath{0.0377\pm0.0005}}		%
\newcommand{\hatcurPPrho}{\ensuremath{0.876^{+0.17}_{-0.24}}}		%

\newcommand{\hatcurPPm}{\ensuremath{1.78^{+0.08}_{-0.05}}}		%
\newcommand{\hatcurPPmlong}{\ensuremath{1.776^{+0.077}_{-0.049}}}	%
\newcommand{\hatcurPPr}{\ensuremath{1.36^{+0.20}_{-0.09}}}		%
\newcommand{\hatcurPPrlong}{\ensuremath{1.363^{+0.195}_{-0.087}}}	%
\newcommand{\hatcurPPmrcorr}{\ensuremath{0.89}}				%
\newcommand{\hatcurXdist}{\ensuremath{320^{+50}_{-40}}}			%

\shortauthors{P\'al et al.}
\shorttitle{HAT-P-7b}

\begin{document}

\title{\hatcur\lowercase{b}: An Extremely Hot Massive Planet Transiting 
	a Bright Star in the Kepler Field}

\author{
	A.~P\'al\altaffilmark{1,2},
	G.~\'A.~Bakos\altaffilmark{1,3},
	G.~Torres\altaffilmark{1},
	R.~W.~Noyes\altaffilmark{1},
	D.~W.~Latham\altaffilmark{1},
	G\'eza~Kov\'acs\altaffilmark{4},
	G.~W.~Marcy\altaffilmark{5},
	D.~A.~Fischer\altaffilmark{6},
	R.~P.~Butler\altaffilmark{7},
	D.~D.~Sasselov\altaffilmark{1},
	B.~Sip\H{o}cz\altaffilmark{2,1},
	G.~A.~Esquerdo\altaffilmark{1},
	G\'abor~Kov\'acs\altaffilmark{1},
	R.~Stefanik\altaffilmark{1},
	J.~L\'az\'ar\altaffilmark{8},
	I.~Papp\altaffilmark{8} \&
	P.~S\'ari\altaffilmark{8}
}
\altaffiltext{1}{Harvard-Smithsonian Center for Astrophysics,
	Cambridge, MA, apal@cfa.harvard.edu}

\altaffiltext{2}{Department of Astronomy,
	E\"otv\"os Lor\'and University, Budapest, Hungary.}

\altaffiltext{3}{NSF Fellow}

\altaffiltext{4}{Konkoly Observatory, Budapest, Hungary}

\altaffiltext{5}{Department of Astronomy, University of California,
	Berkeley, CA}

\altaffiltext{6}{Department of Physics and Astronomy, San Francisco
	State University, San Francisco, CA}

\altaffiltext{7}{Department of Terrestrial Magnetism, Carnegie
	Institute of Washington, DC}

\altaffiltext{8}{Hungarian Astronomical Association, Budapest, 
	Hungary}

\altaffiltext{$\dagger$}{%
	Based in part on observations obtained at the W.~M.~Keck
	Observatory, which is operated by the University of California and
	the California Institute of Technology. Keck time has been in part
	granted by NOAO.
}


\begin{abstract} 

We report on the latest discovery of the HATNet project; a very hot
giant planet orbiting a bright ($V=\hatcurCCtassmvshort$) star with a
small semi-major axis of $a=\hatcurPParel$\,AU\@. Ephemeris for the
system is $P=\hatcurLCP$~days, mid-transit time $E=\hatcurLCT$~(BJD).
Based on the available spectroscopic data on the host star and
photometry of the system, the planet has a mass of
$M_p=\hatcurPPm$\,\mjuplong\ and radius of $R_p=\hatcurPPr$\,\rjuplong.
The parent star is a slightly evolved \hatcurYYspec{} star with
$M_\star=\hatcurYYm$\,\msun{}, $R_\star=\hatcurYYr$\,\rsun,
$\teffstar=\hatcurSMEteff$\,K, and metallicity
$[\mathrm{Fe/H}]=\hatcurSMEzfeh$.  The relatively hot and large host
star, combined with the close orbit of the planet, yield a very high
planetary irradiance of $(4.71^{+1.44}_{-0.05}) \times 10^9$ 
erg cm$^{-2}$ s$^{-1}$, which places the planet near the top of the
pM class of irradiated planets as defined by  \citet{Fortney:07b}.  If
as predicted by \citet{Fortney:07b} the planet re-radiates its
absorbed energy before distributing it to the night side, the day-side
temperature should be about $(2730^{+150}_{-100})$\,K.
Because the host star
is quite bright, measurement of the secondary eclipse should be
feasible for ground-based telescopes, providing a good opportunity to
compare the predictions of current hot Jupiter atmospheric models with
the observations. Moreover, the host star falls in the field of the
upcoming Kepler mission; hence extensive space-borne follow-up,
including not only primary transit and secondary eclipse observations
but also asteroseismology, will be
possible.

\end{abstract}



\keywords{ 
	planetary systems ---
	stars: individual (\hatcur{}, \hatcurCCgsc{}) 
	techniques: spectroscopic
}



\section{Introduction}
\label{sec:introduction}

Transiting planets provide unique information about the nature and
evolution of extrasolar planets because they yield direct measurement
of the radius and (together with radial velocity data) mass of these
objects. Many of these planets are found in quite close-in orbits of
their parent stars, in which case the radiation from the parent star
may be expected to play a major role in controlling the structure and
dynamics of their atmospheres. Transiting planets with very tight
orbits about their parent stars are heated sufficiently that their
thermal emission is measurable, both at times of secondary eclipse and
for some, throughout their orbit \citep[see
e.g.][]{deming05,knutson07,harrington07}. Such observations, together
with observations in different spectral bands of the depth and shape
of primary transits, open up the possibility of detailed understanding
of the composition, structure, and dynamics of ``hot Jupiter''
atmospheres. (See e.g. \citet{Fortney:07b}, \citet{hubeny03},
\citet{burrows06}.

Here
we report on the HATNet project's detection of a \vhj{} that transits a
bright, $V=\hatcurCCtassmvshort$ \hatcurYYspec{} star, with a $2.2$~day
period and semi-major axis of only $0.038$\,AU\@. The parent star
(hereafter denoted \hatcur{}), with $\teffstar=6350$\,K and radius
$1.8$\,\rsun\ has such a high luminosity as to make the surface
temperature of this very close-orbiting planet among the highest of any
known transiting planet.

Because \hatcur{} is relatively bright, this opens up the possibility
of a number of interesting follow-on studies from space and ground.
Notably, the host star is also in the field of view of the forthcoming
Kepler mission. This means that many transits will very likely be
recorded with extremely high photometric precision during the
operational phase of Kepler starting in 2009.  Timing variations in the
transits, if detected, could indicate the presence of other bodies in
the system, such as terrestrial-mass planets or Trojan satellites. For
this reason, preparatory observations of transits and secondary
eclipses in the \hatcur{} system undertaken in advance of the mission
could significantly increase the scientific return from later
observations of the star by Kepler.

The next section (\S{}~\ref{sec:detection}) describes the photometric
detection of the \hatcurb{} planet by the HATNet monitoring system;
\S{}~\ref{sec:followup} outlines both the photometric and spectroscopic
follow-up observations; \S{}~\ref{sec:analysis} explains our analysis
of these data including derivation of stellar and planetary parameters
as well as verification that the signal derives from a transiting
planet rather than a ``false-positive''; and finally,
\S{}~\ref{sec:discussion} discusses the results and implications for
further studies of this intriguing system.


\section{Photometric detection}
\label{sec:detection}

The HATNet telescopes \mbox{HAT-7} and \mbox{HAT-8}
\citep[HATNet;][]{Bakos:02, Bakos:04} observed HATNet field G154,
centered at $\alpha = 19^{\rm h} 12^{\rm m}$, $\delta = +45\arcdeg
00\arcmin$, on a near-nightly basis from 2004 May 27 to 2004 August 6.
Exposures of 5 minutes were obtained at a 5.5-minute cadence whenever
conditions permitted; all in all 5140 exposures were secured, each
yielding photometric measurements for approximately $33,000$ stars in
the field down to $I\sim13.0$. The field was observed in network mode,
exploiting the longitude separation between \mbox{HAT-7}, stationed at
the Smithsonian Astrophysical Observatory's (SAO) Fred Lawrence Whipple
Observatory (FLWO) in Arizona ($\lambda=111\arcdeg$W), and
\mbox{HAT-8}, installed on the rooftop of SAO's Submillimeter Array
(SMA) building atop Mauna Kea, Hawaii ($\lambda=155\arcdeg$W). We note that
each \lc{} obtained by a given instrument was shifted to
have a median value to be the same as catalogue magnitude of the 
appropriate star, allowing to merge \lc{}s acquired by different 
stations and/or detectors.

Following standard frame calibration procedures, astrometry was
performed as described in \citet{Pal:06}, and aperture photometry
results were subjected to External Parameter Decorrelation \citep[EPD,
described briefly in][]{Bakos:07}, and also to the Trend Filtering
Algorithm \citep[TFA;][]{Kovacs:05}. We searched the \lcs{} of field
G154 for box-shaped transit signals using the BLS algorithm of
\citet{Kovacs:02}. A very significant periodic dip in brightness was
detected in the $I\approx\hatcurCCmag$ magnitude star \hatcurCCgsc{}
(also known as \hatcurCCtwomass{}; $\alpha = \hatcurCCra$, $\delta =
\hatcurCCdec$; J2000), with a depth of $\sim\hatcurLCdip$\,mmag, a
period of $P=\hatcurLCPshort$\,days and a relative duration (first to
last contact) of $q\approx0.078$, equivalent to a duration of
$Pq\approx4.1$~hours.

In addition, the star happened to fall in the overlapping area between
fields G154 and G155. Field G155, centered at $\alpha = 19^{\rm h}
48^{\rm m}$, $\delta = +45\arcdeg 00\arcmin$, was also observed over an
extended time in between 2004 July 27 and 2005 September 20 by the
\mbox{HAT-6} (Arizona) and \mbox{HAT-9} (Hawaii) telescopes. We
gathered 1220 and 10260 data-points, respectively (which independently
confirmed the transit), yielding a total number of 16620 data-points.
The combined HATNet \lc{} after the EPD and TFA procedures is plotted
on Figs.~\ref{fig:lc}a,b, superimposed on
these plots is our best fit model (see \S~\ref{sec:analysis}).
We note that TFA was run in signal reconstruction mode, i.e. systematics
were iteratively filtered out from the observed time series assuming that
the underlying signal is a trapeze-shaped transit 
\citep[see][for additional details]{Kovacs:05}.

In addition to having a significant overlap with each other, we note
that fields G154 and G155 both intersect the field of view of the
upcoming Kepler mission \citep{Borucki:06}.


\section{Follow-up observations}
\label{sec:followup}

\subsection{Low-resolution Spectroscopy}

Following the HATNet photometric detection, \hatcur{} (then a transit
{\em candidate}) was observed spectroscopically with the CfA Digital
Speedometer \citep[DS, see][]{Latham:92} at the \flwos{} Tillinghast reflector, in
order to rule out a number of blend scenarios that mimic planetary
transits \citep[e.g.][]{brown03,donovan07}, as well as to characterize
the stellar parameters, such as surface gravity, effective temperature,
and rotation. Four spectra were obtained over an interval of 29 days. 
These observations cover $45$\,\AA{} in a single echelle order centered
at $5187$\,~\AA{}, and have a resolving power of $\lambda/\Delta\lambda
\approx 35,\!000$.  Radial velocities were derived by
cross-correlation, and have a typical precision of $1$\,\kms. Using
these measurements, we have ruled out an unblended companion of stellar
mass (e.g.~an M dwarf orbiting an F dwarf), since the radial velocities
did not show any variation within the uncertainties. The mean
heliocentric radial velocity of \hatcur{} was measured to be
$-11$\,~\kms.  Based on an analysis similar to that described in
\cite{Torres:02}, the DS spectra indicated that the host star is a
slightly evolved dwarf with $\logg = 3.5$ (cgs), $\teffstar = 6250$\,K
and $\vsini \approx 6\,\kms$.

\begin{figure}[!ht]
\plotone{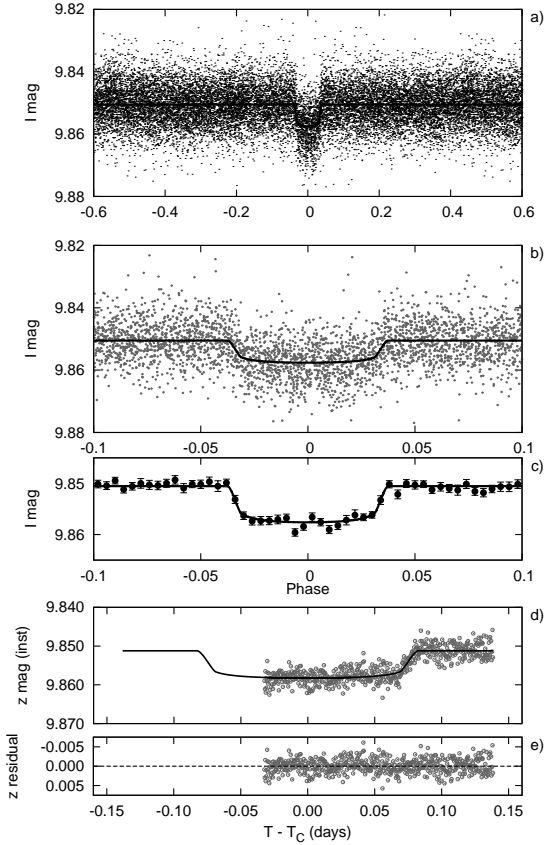}
\caption{
	(a) The complete \lc{} of \hatcur{} with all of the 16620 points, 
	unbinned instrumental \band{I}
	photometry obtained with four telescopes of HATNet (see text for
	details), and folded with the period of $P = \hatcurLCPprec$~days
	(the result of a joint fit to all available data,
	\S~\ref{sec:jointfit}). The superimposed curve shows the best model
	fit using quadratic limb darkening.
	(b) Same as a), with the transit zoomed-in (3150 data points are shown).
	(c) Same as b), with the points binned with a bin size of 0.004 in phase.
	(d) Unbinned instrumental Sloan \band{z} partial transit photometry
	acquired by the KeplerCam at the \flwof{} telescope on 2007
	November 2; superimposed is the best-fit transit model \lc{}.
	(e) The difference between the KeplerCam observation and model
	(same vertical scale as in panel d).
\label{fig:lc}}
\end{figure}

\subsection{High resolution spectroscopy}

For the characterization of the radial velocity variations and for the
more precise determination of the stellar parameters, we obtained 8
exposures with an iodine cell, plus one iodine-free template, using the
HIRES instrument \citep{Vogt:94} on the Keck~I telescope, Hawaii,
between 2007 August 24 and 2007 September 1. The width of the
spectrometer slit was $0\farcs86$ resulting a resolving power of
$\lambda/\Delta\lambda \approx 55,\!000$, while the wavelength coverage
was $\sim3800-8000$\,\AA\@. The iodine gas absorption cell was used to
superimpose a dense forest of $\mathrm{I}_2$ lines on the stellar
spectrum and establish an accurate wavelength fiducial
\citep[see][]{Marcy:92}. Relative radial velocities in the Solar System
barycentric frame were derived as described by \cite{Butler:96},
incorporating full modeling of the spatial and temporal variations of
the instrumental profile. The final radial velocity data and their
errors are listed in Table~\ref{tab:rvs}. The folded data, with our
best fit (see \S~\ref{sec:jointfit}) superimposed, are plotted in
Fig.~\ref{fig:rvbis}a.

\subsection{Photometric follow-up observations}

Partial photometric coverage of a transit event of \hatcur{} was
carried out in the Sloan \band{z} with the KeplerCam CCD on the
\mbox{1.2 m} telescope at FLWO, on 2007 November 2. The total number of
frames taken from \hatcur{} was 514 with cadence of 28 seconds. During
the reduction of the KeplerCam data, we used the following method.
After bias and flat calibration of the images, an astrometric
transformation (in the form of first order polynomials)  between the
$\sim450$ brightest stars and the 2MASS catalog was derived, as
described in \citet{Pal:06}, yielding a residual of $\sim1/4$ pixel.
Aperture photometry was then performed
using a series of apertures of with the radius of 4, 6 and 8 pixels
in fixed positions calculated from this
solution and the actual 2MASS positions. The instrumental magnitude
transformation was obtained using $\sim350$ stars on a frame taken near
culmination of the field. The transformation fit was initially weighted
by the estimated photon- and background-noise error of each star, then
the procedure was repeated weighting by the inverse variance of the
\lcs{}. From the set of apertures we have chosen the aperture for which
the out-of-transit (OOT) rms of \hatcur{} was the smallest; the radius
of this aperture is 6 pixels. This \lc{}
was then de-correlated against trends using the OOT sections, which
yielded a light curve with an overall rms of 1.9mmag at a cadence of
one frame per 28 seconds. This is a bit larger than the expected rms of
1.5mmag, derived from the photon noise (1.2mmag) and scintillation noise
-- which has an expected amplitude of 0.8mmag, based on the observational
conditions and the calculations of \citet{Young:67} -- 
possibly due to unresolved trends and other noise sources. The resulting
\lc{} is shown in Fig.~\ref{fig:lc}d.


\section{Analysis}
\label{sec:analysis}

The analysis of the available data was done in five steps. First, an
independent analysis was performed on the HATNet, the radial velocity
(RV) and the high precision photometric follow-up (FU) data,
respectively. Analysis of the HATNet data yielded an initial value for
the orbital period and transit epoch. The initial period and epoch were
used to fold the RV's, and phase them with respect to the predicted
transit time for a circular orbit. The HATNet and the RV epochs
together yield a more accurate period, since the time difference
between the discovery \lc{} and the RV follow-up is fairly long; more
than 3 years.  Using this refined period, we can extrapolate to the
expected center of the KeplerCam partial transit, and therefore obtain
a fit for the two remaining key parameters describing the light curve:
$a/R_\star$ where $a$ is the semi-major axis for a circular orbit, and
the impact parameter $b\equiv(a/R_\star)\cos i$, where $i$ is the
inclination of the orbit.

Second, using as starting points the initial values as derived above,
we performed a joint fit of the HATNet, RV and FU data, i.e.~fitting
\emph{all} of the parameters simultaneously. The reason for such a
joint fit is that the three separate data-sets and the fitted
parameters are intertwined. For example, the epoch (depending partly on
the RV fit) has a relatively large error, affecting the extrapolation
of the transit center to the KeplerCam follow-up.

In all of the above procedures, we used the downhill simplex method to
search for the best fit values and the method of refitting to synthetic
data sets to find out the error of the adjusted parameters. The latter
method yields a Monte-Carlo set of the \emph{a posteriori} distribution
of the fit parameters.

The third step of the analysis was the derivation of the stellar
parameters, based on the spectroscopic analysis of the host star (high
resolution spectroscopy using Keck/HIRES), and the physical modeling of
the stellar evolution, based on existing isochrone models. As the
fourth step, we then combined the results of the joint fit and stellar
parameter determination to determine the planetary and orbital
parameters of the \hatcurb{} system.  Finally, we have done a blend
analysis to confirm that the orbiting companion is a \hj{}. In the
following subsections we give a detailed description of the above main
steps of the analysis.

\begin{deluxetable}{lrrc}
\tablewidth{0pc}
\tablecaption{Relative radial velocity measurements 
of \hatcur{}\label{tab:rvs}}
\tablehead{
	\colhead{BJD} & 
	\colhead{RV} & 
	\colhead{\ensuremath{\sigma_{\rm RV}}
}\\
\colhead{
	\hbox{~~~~(2,454,000$+$)~~~~}} & 
	\colhead{(\ms)} & 
	\colhead{(\ms)}
}
\startdata
336.73958  \dotfill &  $+124.40$ &   $            1.63  $ & \\
336.85366  \dotfill &  $+ 73.33$ &   $            1.48  $ & \\     
337.76211  \dotfill &  $-223.89$ &   $            1.60  $ & \\     
338.77439  \dotfill &  $+166.71$ &   $            1.39  $ & \\     
338.85455  \dotfill &  $+144.67$ &   $            1.42  $ & \\     
339.89886  \dotfill &  $-241.02$ &   $            1.46  $ & \\     
343.83180  \dotfill &  $-145.42$ &   $            1.66  $ & \\     
344.98804  \dotfill &  $+101.05$ &   $            1.91  $ & 
\enddata
\end{deluxetable}

\subsection{Independent fits}
\label{sec:independentfit}

For the independent fit procedure, we first analyzed the HATNet \lcs{},
as observed by the \mbox{HAT-6}, \mbox{HAT-7}, \mbox{HAT-8} and
\mbox{HAT-9} telescopes. Using the initial period and transit length
from the BLS analysis, we fitted a model to the 214 cycles of
observations spanned by all the HATNet data.  Although at this stage we
were interested only in the epoch and period, we have used the transit
\lc{} model with the assumption quadratic limb darkening, where the
flux decrease was calculated using the models provided by
\citet{Mandel:02}. In principle, fitting the epoch and period as two
independent variables is equivalent to fitting the time instant of the
centers of the first and last observed individual transits,
$T_{\mathrm{c,first}}$ and $T_{\mathrm{c,last}}$, with a constraint
that all intermediate transits are regularly spaced with period $P$.
Note that this fit takes into account {\em all}\/ transits that
occurred during the HATNet observations, even though it is described
only by $T_{\mathrm{c,first}}$ and $T_{\mathrm{c,last}}$.
The fit yielded 
$T_{\mathrm{c,first}}=2453153.0924\pm0.0021$ (BJD) and
$T_{\mathrm{c,last}} =2453624.9044\pm0.0023$ (BJD). 
%
The  period 
derived from the 
$T_{\mathrm{c,first}}$
and $T_{\mathrm{c,last}}$ epochs  was
$P^{(1)}=2.20480\pm0.00049$~days.
Using these values, we found that there were 326 cycles between 
$T_{\mathrm{c,last}}$ 
and the end of the RV campaign. The epoch
extrapolated to the approximate time of RV measurements was
$T_{\mathrm{c,RV}}=2454343.646\pm0.008$ (BJD). Note that the error in
$T_{\mathrm{c,RV}}$ is 
much smaller than the period itself
($\sim2.2$\,days), 
so there is no ambiguity in the number of elapsed
cycles when folding the periodic signal. 

We then analyzed the radial velocity data in the following way. We
defined the $N_{\rm tr}\equiv0$ transit as that being closest to the
end of the radial velocity measurements. This means that the first
transit observed by HATNet (at $T_{\mathrm{c,first}}$) was the $N_{\rm
tr,first}=-540$ event. Given the short period, we assumed that the
orbit has been circularized \citep{hut81} (later verified; see below). 
The orbital fit is linear if we choose the radial velocity zero-point
$\gamma$ and the amplitudes $A$ and $B$ as adjusted values, namely:
\begin{equation}
	v(t) = \gamma + A\cos\left(\frac{2\pi}{P}(t-t_0)\right) + 
		B\sin\left(\frac{2\pi}{P}(t-t_0)\right),
\end{equation}
where $t_0$ is an arbitrary time instant (chosen to be $t_0=2454342.6$
BJD), $K\equiv\sqrt{A^2+B^2}$ is the semi-amplitude of the RV
variations, and $P$ is the initial period $P^{(1)}$ taken from the
previous independent HATNet fit. The actual epoch can be derived from
the above equation since for circular orbits transit center occurs when
the RV curve has the most negative slope.
Using the equations above, we derived the initial epoch of the $N_{\rm
tr} = 0$ transit center to be $T_c=2454343.6462\pm0.0042\equiv
T^{(1)}_{\mathrm{c},0}$ (BJD). We also performed a more general
(non-linear) fit to the RV in which we let the eccentricity float. 
This fit yielded an eccentricity consistent with zero, namely
$e\cos\omega=-0.003\pm0.007$ and $e\sin\omega= 0.000\pm0.010$.
Therefore, we adopt a circular orbit in the further analysis.

Combining the RV epoch $T^{(1)}_{\mathrm{c},0}$ with the first epoch
observed by HATNet ($T_{\mathrm{c,first}}$), we obtained a somewhat
refined period, $P^{(2)}=2.204732\pm0.000016$~days. This was fed back
into phasing the RV data, and we performed the RV fit again to the
parameters $\gamma$, $A$ and $B$. The fit yielded
$\gamma=-37.0\pm1.5$\,\ms, $K\equiv\sqrt{A^2+B^2}= 213.4\pm2.0$\,\ms
and $T^{(2)}_{\mathrm{c},0}=2454343.6470\pm0.0042$ (BJD). This epoch
was used to further refine the period to get
$P^{(3)}=2.204731\pm0.000016$\,d, where the error calculation assumes
that $T_{\mathrm{c},0}$ and $T_{\mathrm{c},-540}$ are uncorrelated. At
this point we stopped the above iterative procedure of refining the
epoch and period; instead a final refinement of epoch and period was
obtained through performing a joint fit, (as described later in
\S~\ref{sec:jointfit}). We note that in order to get a reduced
chi-square value near unity for the radial velocity fit, it was
necessary to quadratically increase the noise component with an
amplitude of $3.8$~\ms, which is well within the range of stellar
jitter observed for late F stars; see \cite{Butler:06}.

Using the improved period $P^{(3)}$ and the epoch $T_{\mathrm{c},0}$,
we extrapolated to the center of KeplerCam follow-up transit ($N_{\rm
tr}=29$).  Since the follow-up observation only recorded a partial
event (see Fig.~\ref{fig:lc}d), this extrapolation was necessary to
improve the \lc{} modeling. For this, we have used a quadratic
limb-darkening approximation, based on the formalism of
\citet{Mandel:02}. The limb-darkening coefficients were based on the
results of the SME analysis (notably, \teffstar; see
\S~\ref{sec:stellarparameters} for further details), which yielded
$\gamma_1^{(z)}=0.1329$ and $\gamma_2^{(z)}=0.3738$.  Using these
values and the extrapolated time of the transit center, we adjusted the
\lc{} parameters: the relative radius of the planet $p=R_p/R_\star$,
the square of the impact parameter $b^2$ and the quantity
$\zeta/R_\star=(a/R_\star)(2\pi/P)(1-b^2)^{-1/2}$ as independent
parameters \citep[see][for the choice of parameters]{Bakos:07b}. The
result of the fit was $p=0.0762\pm0.0012$, $b^2=0.205\pm0.144$ and
$\zeta/R_\star=13.60\pm0.83$\,day$^{-1}$, where the uncertainty of the
transit center time due to the relatively high error in the transit
epoch $T_{\mathrm{c},0}$ was also taken into account in the error
estimates.

\begin{figure} 
\plotone{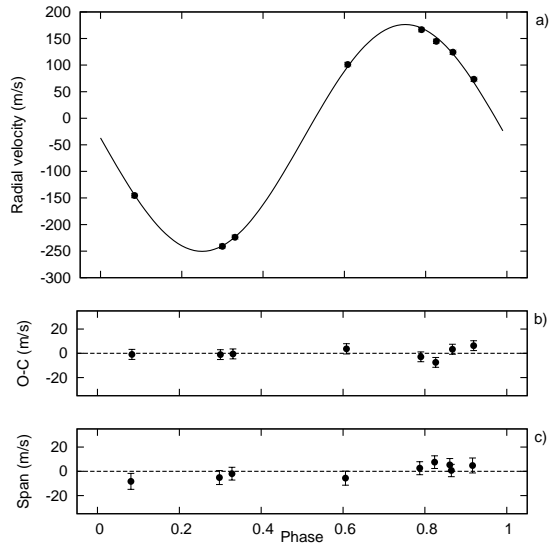}
\caption{
(a) Radial-velocity measurements from Keck for \hatcur{}, along with an
orbital fit, shown as a function of orbital phase, using our best fit
as period (see \S~\ref{sec:jointfit}). The center-of-mass velocity has
been subtracted.
(b) Phased residuals after subtracting the orbital fit 
(also see \S~\ref{sec:jointfit}). The rms 
variation of the residuals is about $3.8$\,\ms. 
(c) Bisector spans (BS) for the 8 Keck spectra plus the single template
spectrum, computed as described in the text.  The mean value has been
subtracted. Due to the relatively small errors comparing to the RV
amplitude, the vertical scale on the (b) and (c) panels differ from the
scale used on the top panel. 
\label{fig:rvbis}}
\end{figure}

\subsection{Joint fit}
\label{sec:jointfit}

The results of the individual fits described above provide the starting
values for a joint fit, i.e.~a simultaneous fit to all of the available
HATNet, radial velocity and the partial follow-up \lc{} data.  The
adjusted parameters were $T_{\mathrm{c},-540}$, the time of first
transit center in the HATNet campaign, $m$, the out-of-transit
magnitude of the HATNet \lc{} in \band{I} and the previously defined
parameters of $\gamma$, $A$, $B$, $p$, $b^2$ and $\zeta/R_\star$.  We
note that in this joint fit {\em all}\/ of the transits in the HATNet
\lc{} have been adjusted simultaneously, tied together by the
constraint of assuming a strictly periodic signal; the shape of all
these transits were characterized by $p$, $b^2$ and $\zeta/R_\star$
(and the limb-darkening coefficients) while the distinct transit center
time instants were interpolated using $T_{\mathrm{c},-540}
=T_{\mathrm{c,first}}$ and $A$, $B$ via the RV fit. For initial values
we used the results of the independent fits (\secr{independentfit}).
The error estimation based on method refitting to synthetic data sets
gives the distribution of the adjusted values, and moreover, this
distribution can be used directly as an input for a Monte-Carlo
parameter determination for stellar evolution modeling, as described
later (\S~\ref{sec:stellarparameters}).

Final results of the joint fit were:
$T_{\mathrm{c},-540}=2453153.0924\pm0.0015$~(BJD), 
$m=9.85053\pm0.00015$\,mag,
$\gamma=\hatcurRVgamma$\,\ms,
$A=33.8\pm0.9$\,\ms,
$B=210.7\pm1.9$\,\ms,
$p=0.0763\pm0.0010$, 
$b^2=0.135_{-0.116}^{+0.149}$ and 
$\zeta/R_\star=13.34\pm0.23$~$\mathrm{day^{-1}}$. 
Using the distribution of these parameters, it is
straightforward to obtain the values and the errors of the additional 
parameters derived from the joint derived fit, namely 
$T_{\mathrm{c},0}$, $a/R_\star$, $K$ and $P$. 
All final fit parameters are listed in Table~\ref{tab:parameters}. 

\begin{deluxetable}{lcl}
\tablewidth{0pc}
\tablecaption{Stellar parameters for \hatcur{} \label{tab:stellar}}
\tablehead{\colhead{~~~~~~Parameter~~~~~~} & \colhead{Value} & \colhead{Source}}
\startdata
$\teffstar$ (K)\dotfill		&  \hatcurSMEteff		& SME\tablenotemark{a} \\
$[\mathrm{Fe/H}]$\dotfill		&  \hatcurSMEzfeh		& SME \\
$v \sin i$ (\kms)\dotfill		&  \hatcurSMEvsin		& SME \\
$M_\star$ ($M_{\sun}$)\dotfill  &  \hatcurYYm			& Y$^2$+LC+SME\tablenotemark{b} \\
$R_\star$ ($R_{\sun}$)\dotfill  &  \hatcurYYr			& Y$^2$+LC+SME \\
$\log g_\star$ (cgs)\dotfill    &  \hatcurYYlogg		& Y$^2$+LC+SME\\
$L_\star$ ($L_{\sun}$)\dotfill  &  \hatcurYYlum			& Y$^2$+LC+SME \\
$M_V$ (mag)\dotfill				&  \hatcurYYmv   		& Y$^2$+LC+SME \\
Age (Gyr)\dotfill				&  \hatcurYYage			& Y$^2$+LC+SME \\
Distance (pc)\dotfill			&  \hatcurXdist			& Y$^2$+LC+SME
\enddata
\tablenotetext{a}{SME = `Spectroscopy Made Easy' package for analysis
of high-resolution spectra \cite{Valenti:96}. See text.}
\tablenotetext{b}{Y$^2$+LC+SME = Yale-Yonsei isochrones \citep{Yi:01},
\lc{} parameters, and SME results.}
\end{deluxetable}

\subsection{Stellar parameters}
\label{sec:stellarparameters}

The results of the joint fit enable us to refine the parameters of the
star. First, the iodine-free template spectrum from Keck was used for
an initial determination of the atmospheric parameters. Spectral
synthesis modeling was carried out using the SME software
\citep{Valenti:96}, with wavelength ranges and atomic line data as
described by \citet{Valenti:05}. We obtained the following initial
values: effective temperature $\hatcurSMEteff$\,K, surface gravity
$\log g_\star = \hatcurSMElogg$ (cgs), iron abundance
$\mathrm{[Fe/H]}=\hatcurSMEzfeh$, and projected rotational velocity
$v\sin i=\hatcurSMEvsin$\,\kms.  The rotational velocity is slightly
smaller than the value given by the DS measurements. The temperature
and surface gravity correspond to a slightly evolved \hatcurYYspec{}
star. The uncertainties quoted here and in the remaining of this
discussion are twice the statistical uncertainties for the values given
by the SME analysis. This reflects our attempt, based on prior
experience, to incorporate systematic errors (e.g. \cite{noyes:08};
see also \cite{Valenti:05}). 
Note that the previously
discussed limb darkening coefficients, $\gamma_1^{(z)}$,
$\gamma_2^{(z)}$, $\gamma_1^{(I)}$ and $\gamma_2^{(I)}$ have been taken
from the tables of \cite{Claret:04} by interpolation to the
above-mentioned SME values for $\teffstar$, $\log g_\star$, and
$\mathrm{[Fe/H]}$.

As described by \cite{Sozzetti:07}, $a/R_\star$ is a better luminosity
indicator than the spectroscopic value of $\log g_\star$ since the
variation of stellar surface gravity has a subtle effect on the line
profiles. Therefore, we used the values of $\teffstar$ and
$\mathrm{[Fe/H]}$ from the initial SME analysis, together with the
distribution of $a/R_\star$ to estimate the stellar properties from
comparison with the Yonsei-Yale (Y$^2$) stellar evolution models by
\cite{Yi:01}.  Since a Monte-Carlo set for $a/R_\star$ values has been
derived during the joint fit, we performed the stellar parameter
determination as follows. For a selected value of $a/R_\star$, two
Gaussian random values were drawn for $\teffstar$ and $\mathrm{[Fe/H]}$
with the mean and standard deviation as given by SME (with formal SME
uncertainties doubled as indicated above).Using these
three values, we searched the nearest isochrone and the corresponding
mass by using the interpolator provided by \citet{Demarque:04}. 
Repeating this procedure for values of $a/R_\star$, $\teffstar$,
$\mathrm{[Fe/H]}$, the set of the \emph{a posteriori} distribution of
the stellar parameters was obtained, including the mass, radius, age,
luminosity and color (in multiple bands).  The age determined in this
way is $2.2$~Gy with a statistical uncertainty of $\pm 0.3$~Gy;
however, the uncertainty in the theoretical isochrone ages is about
1.0~Gy. Since the corresponding value for the surface gravity of the
star, $\log g_\star=\hatcurYYlogg$ (cgs),is well within 1-$\sigma$ of
the value determined by the SME analysis, we accept the values from the
joint fit as the final stellar parameters. These parameters are
summarized in Table~\ref{tab:stellar}.

We note that the Yonsei-Yale isochrones contain the absolute magnitudes and
colors for
different photometric bands from $U$ up to $M$, providing an easy
comparison of the estimated and the observed colors. Using these data,
we determined the $V-I$ and $J-K$ colors of the best fitted stellar
model:
$(V-I)_{\rm YY}=0.54\pm0.02$ and 
$(J-K)_{\rm YY}=0.27\pm0.02$.
Since the colors for the infrared bands provided by \citet{Yi:01} and
\citet{Demarque:04} are given in the ESO photometric standard system,
for the comparison with catalog data, we converted the infrared color
$(J-K)_{\rm YY}$ to the 2MASS system $(J-K_S)$ using the
transformations given by \citet{Carpenter:01}. The color of the best
fit stellar model was $(J-K_S)_{\rm YY}=0.25\pm0.03$, which is in
fairly good agreement with the actual 2MASS color of \hatcur{}:
$(J-K_S)=0.22\pm0.04$. We have also compared the $(V-I)_{\rm YY}$ color
of the best fit model to the catalog data, and found that although
\hatcur{} has a low galactic latitude, $b_{\rm II}=13\fdg8$, the model
color agrees well with the observed TASS color of $(V-I)_{\rm
TASS}=\hatcurCCtassvi$ \citep[see][]{Droege:06}. Hence, the star is not
affected by the interstellar reddening within the errors, since
$E(V-I)\equiv(V-I)_{\rm TASS}-(V-I)_{\rm YY}=0.06\pm0.07$.  For
estimating the distance of \hatcur, we used the absolute magnitude
$M_V=\hatcurYYmv$ (resulting from the isochrone analysis, see also 
Table~\ref{tab:stellar}) and the $V_{\rm TASS}=\hatcurCCtassmv\pm0.06$
observed magnitude. These two yield a distance modulus of $V_{\rm
TASS}-M_V=7.51\pm0.28$, i.e.~distance of $d=\hatcurXdist$\,pc.

\subsection{Planetary and orbital parameters}

The determination of the stellar properties was followed by the
characterization of the planet itself. Since Monte-Carlo distributions
were derived for both the \lc{} and the stellar parameters, the final
planetary and orbital data were also obtained by the statistical
analysis of the \emph{a posteriori} distribution of the appropriate
combination of these two Monte-Carlo data sets.  We found that the mass
of the planet is $M_p=\hatcurPPmlong$\,\mjup, the radius is
$R_p=\hatcurPPrlong$\,\rjup\ and its density is
$\rho_p=\hatcurPPrho$\,\gcmc.  We note that in the case of binary
systems with large mass and radius ratios (such as the one here) there
is a strong correlation between $M_p$ and $R_p$ \citep[see
e.g.][]{Beatty:07}. This correlation is also exhibited here with
$C(M_p,R_p)=\hatcurPPmrcorr$.  The final planetary parameters are also
summarized at the bottom of Table~\ref{tab:parameters}.

Due to the way we derived the period, i.e.
$P=(T_{\mathrm{c},0}-T_{\mathrm{c},-540})/540$, one can expect a large
correlation between the epochs $T_{\mathrm{c},0}$,
$T_{\mathrm{c},-540}$ and the period itself. Indeed,
$C(T_{\mathrm{c},-540},P)=-0.783$ and $C(T_{\mathrm{c},0},P)=0.704$,
while the correlation between the two epochs is negligible;
$C(T_{\mathrm{c},-540},T_{\mathrm{c},0})=-0.111$. It is easy to show
that if the signs of the correlations between two epochs $T_{\rm A}$
and $T_{\rm B}$ (in our case $T_{\mathrm{c},0}$ and
$T_{\mathrm{c},-540}$) and the period are different, respectively, then
there exists an optimal epoch $E$, which has the smallest error among
all of the interpolated epochs. We note that $E$ will be such that it
also exhibits the smallest correlation with the period. If
$\sigma(T_{\rm A})$ and $\sigma(T_{\rm B})$ are the respective
uncorrelated errors of the two epochs, then
\begin{equation}
E=\left[\frac{T_{\rm A}\sigma(T_{\rm B})^2+T_{\rm B}\sigma(T_{\rm A})^2}
	{\sigma(T_{\rm B})^2+\sigma(T_{\rm A})^2}\right]
\end{equation}
where square brackets denote the time of the transit event nearest to
the time instance $t$. In the case of \hatcurb{}, $T_{\rm A}\equiv
T_{\mathrm{c},-540}$ and $T_{\rm B}\equiv T_{\mathrm{c},0}$, the
corresponding epoch is the event $N_{\rm tr}=-251$ at $E\equiv
T_{\mathrm{c},-251}=\hatcurLCT$ (BJD).  The final ephemeris and
planetary parameters are summarized in Table~\ref{tab:parameters}.

\begin{deluxetable}{lc}
\tablewidth{0pc}
\tablecaption{Orbital and planetary parameters\label{tab:parameters}}
\tablehead{\colhead{~~~~~~~~~~~~~~Parameter~~~~~~~~~~~~~~} & \colhead{Value}}
\startdata

\sidehead{\Lc{} parameters}
~~~$P$ (days)			\dotfill        		& $\hatcurLCP$ 		\\
~~~$E$ (${\rm BJD}-2,\!400,\!000$)	\dotfill		& $\hatcurLCMT$		\\
~~~$T_{14}$ (days)\tablenotemark{a} \dotfill	& $\hatcurLCdur$		\\
~~~$T_{12} = T_{34}$ (days)\tablenotemark{a} \dotfill	& $\hatcurLCingdur$	\\
~~~$a/R_\star$			\dotfill               	& $\hatcurPPar$		\\
~~~$R_p/R_\star$		\dotfill              	& $\hatcurLCrprstar$	\\
~~~$b \equiv a \cos i/R_\star$	\dotfill		& $\hatcurLCimp$		\\
~~~$i$ (deg)			\dotfill               	& $\hatcurPPi$ \phn 	\\

\sidehead{Spectroscopic parameters}
~~~$K$ (\ms)			\dotfill               	& $\hatcurRVK$		\\
~~~$\gamma$ (\kms)		\dotfill 	        	& $\hatcurRVgamma$	\\
~~~$e$				\dotfill					& $0$ (adopted)		\\

\sidehead{Planetary parameters}
~~~$M_p$ ($\mjup$)	\dotfill					& $\hatcurPPmlong$	\\
~~~$R_p$ ($\rjup$)	\dotfill					& $\hatcurPPrlong$	\\
~~~$C(M_p,R_p)$		\dotfill					& $\hatcurPPmrcorr$	\\
~~~$\rho_p$ (\gcmc)	\dotfill					& $\hatcurPPrho$	\\
~~~$a$ (AU)			\dotfill                	& $\hatcurPParel$	\\
~~~$\log g_p$ (cgs)		\dotfill				& $\hatcurPPlogg$	\\
\enddata
\tablenotetext{a}{\ensuremath{T_{14}}: total transit duration, time
between first to last contact; \ensuremath{T_{12}=T_{34}}:
ingress/egress time, time between first and second, or third and fourth
contact.} 
\end{deluxetable}

\subsection{Excluding blend scenarios}

Following \cite{Torres:07}, we explored the possibility that the
measured radial velocities are not real, but instead caused by
distortions in the spectral line profiles due to contamination from a
nearby unresolved eclipsing binary.  In that case the ``bisector span''
of the average spectral line should vary periodically with amplitude
and phase similar to the measured velocities themselves
\citep{Queloz:01,Mandushev:05}. We cross-correlated each Keck spectrum
against a synthetic template matching the properties of the star
(i.e.~based on the SME results, see \S~\ref{sec:stellarparameters}),
and averaged the correlation functions over all orders blueward of the
region affected by the iodine lines. From this representation of the
average spectral line profile we computed the mean bisectors, and as a
measure of the line asymmetry we computed the ``bisector spans'' as the
velocity difference between points selected near the top and bottom of
the mean bisectors \citep{Torres:05}. If the velocities were the result
of a blend with an eclipsing binary, we would expect the line bisectors
to vary in phase with the photometric period with an amplitude similar
to that of the velocities. Instead, we detect no variation in excess of
the measurement uncertainties (see Fig.~\ref{fig:rvbis}c). 
We have also tested the significance of the correlation between the
radial velocity and the bisector variations. 
Therefore, we conclude
that the velocity variations are real and that the star is orbited by a
Jovian planet. We note here that the mean bisector span ratio relative
to the radial velocity amplitude is the smallest ($\sim 0.026$) among
all the HATNet planets, indicating an exceptionally high confidence
that the RV signal is not due to a blend with an eclipsing binary
companion. 


\section{Discussion}  
\label{sec:discussion}

With orbital period of only 2.2 days, implying a semi-major axis of only
$0.038$\,AU, \hatcurb{}, is a ``\vhj'', a name frequently given to
giant planets with orbital period less than 3 days
\citep[see][]{Bouchy:04}.  However, what really matters is the
incident flux from the parent star, which for \hatcurb{} is
exceptionally high because (a) the large radius of the parent star
($1.84$\,\rsun) combined with the small semi-major axis yields a very
small geometrical factor $a/R_\star=4.35$ (whose inverse describes the
stellar flux impinging on the planet); and (b) its parent star has
relatively high effective temperature: $\teffstar=6350$\,K. 
The flux impinging on the planet is $4.7 \times 10^9$ erg cm$^{-2}$s$^{-1}$. 
This places it among the most highly irradiated members of the ``pM'' 
class of planets,
as discussed by \citet{Fortney:07b}.  

For a pM class planet, \citet{Fortney:07b} argue that the incident
radiation is absorbed by TiO and VO molecules in a hot stratosphere,
and almost immediately re-radiated, rather than being partially advected
to the night side of the planet, as would be the case if the radiation
were absorbed deeper down in the cooler atmosphere of a ``pL'' planet,
which would not have
gaseous TiO and VO in its atmosphere. 

For a pM planet, the dayside temperature would be given by 
\begin{equation}
	T_p=\teffstar\left(\frac{R_\star} {a}\right)^{1/2} \lbrack
	f(1-A_B)\rbrack^{1/4}.
\end{equation}
where $A_{\rm B}$ is the Bond albedo, which we assume to be essentially
zero \citep[see, e.g.][]{Rowe:07}, and  the redistribution factor 
$f = 2/3$ \citep[see][]{Lopez:07}. 
For \hatcur{} this yields a dayside temperature of
$(2730^{+150}_{-100})$\,K. 
If at the other extreme, contrary to the theory of \citet{Fortney:07b},
the incident energy absorbed by \hatcurb{} were to be distributed
isotropically over the planet's surface before being re-emitted, as is
predicted for ``pL'' planets too cool to have significant TiO and VO
in their atmospheres, then $f = 1/4$ in Equation 3, and the day-side
temperature would be $(2140^{+110}_{-60})$\,K.  
Thus, detection and measurement of the emission from the dayside of
\hatcurb{} would yield an important test of the theory of
\citet{Fortney:07b}.

\citet{Lopez:07} and \citet{Fortney:07b} noted that the thermal
emission in the optical and near infrared from \vhj{}s should be
detectable from the ground. \citet{Lopez:07} identified OGLE-TR-56b,
the only other known very hot jupiter with predicted dayside
temperature as high as that of \hatcurb{}, as a particularly promising 
candidates.  However, since
\hatcur{} is 6 magnitudes brighter than OGLE-TR-56, it is a far better
candidate for secondary eclipse studies from the ground. For
\hatcur{}, the brightness decrease in \band{z'} at secondary
eclipse should be
$\sim0.2~{\rm mmag}$ or $\sim0.04~{\rm mmag}$ 
in the respective cases
of zero (pM-type) or complete (pL-type) planet-wide re-distribution of 
irradiated stellar flux. This implies that in the former case the
eclipse could be detected at the 2-sigma level with a 1.2-m telescope
with a \band{z'} integration of 4 hours during eclipse (that is, the
total eclipse duration), plus 4 more hours of out-of-eclipse
observation. 
We plan to carry out such observations in the near future with the
KeplerCam on \flwof{} and with MMT~6.5~m (also at FLWO) telescopes.

Recently \citet{Pollaco:07} reported the discovery of the transiting
exoplanet WASP-3b, located 0.0317 AU from an F star with
$\teffstar=6350$\,K and radius $\rstar=1.31\rsun$; this indicates a
very high radiative flux at the planet of $2.5 \times 10^9$ erg
cm$^{-2}$s$^{-1}$, comparable to but about 25\% smaller than the
radiative flux at \hatcurb{}. For the same reasons as for the case of
\hatcurb{} discussed above, this almost certainly will be a pM-class
planet, so that its dayside temperature should be about 2530K,
as indicated by Equation 3 with $f = 2/3$ and $A_B =
0$. This planet should also be detectable in visible or near infrared
radiation. In fact, if it is a pM planet, the depth of secondary
eclipse as seen in \band{z'} should be similar to that for \hatcurb{},
namely $\sim0.2~{\rm mmag}$: although with its lower dayside
temperature  the surface brightness of WASP-3b
should be smaller than for \hatcurb{}, the geometrical factor $(\rpl /
\rstar)^2$ is larger, because WASP-3 is a less evolved star.

\hatcurb{} has an additional important property in that it is in the
field of view of one of the detectors on the forthcoming Kepler mission
\citep{Borucki:06}, currently scheduled for launch in early 2009 to
monitor stars for photometric signals from transiting planets and other
forms of stellar variability. Hence, \hatcur{} should be subjected to
intense and regular photometric observations for several years starting
in 2009. This will permit extremely high precision \lcs{} of the
primary transit, the secondary transit, and perhaps also the variation
of thermal and reflected light emission from the planet over the course
of its orbit as described above.  In addition, it will be possible to
search intensively for transit timing variations over the several year
lifetime of the mission.  Such observations could yield important
information about orbital variations due to other companions in the
system including terrestrial-mass or smaller planets or Trojans, or to
other effects such as precession of the orbital plane or the orbital
line of nodes. To maximize the scientific return from Kepler in this
regard, it would be very useful to obtain high precision \lcs{} both
from the ground, and from space instruments such as Epoxi and MOST, in
the very near term for later comparison with Kepler observations.

Asteroseismology of \hatcur{} using Kepler could yield an independent
measurement of the star's mean density from the ``large separation''
$\Delta\nu$ between p-modes of the same low angular degree $\ell$
\citep[e.g.][]{Stello:07}. This would provide a check on the mean
density of the star completely independent of that derived from the
transit measurements of the $a/R_\star$ parameter.  Equivalently, the
asteroseismology measurements would yield an independent measurement of
the stellar radius, and hence a more precise value for the radius of
the planet \hatcurb{}. Such a rich combined data set could have
additional benefits for understanding stellar structure and evolution.


\acknowledgements 

Operation of the HATNet project is funded in part by NASA grant
NNG04GN74G. Work by G.\'A.B.~was supported by NASA through Hubble
Fellowship Grant HST-HF-01170.01-A and by the Postdoctoral Fellowship
of the NSF Astronomy and Astrophysics Program. We acknowledge partial
support also from the Kepler Mission under NASA Cooperative Agreement
NCC2-1390 (D.W.L., PI). G.T.~acknowledges partial support from NASA
under grant NNG04LG89G, G.K.~thanks the Hungarian Scientific Research
Foundation (OTKA) for support through grant K-60750. A.P.~would like to
thank the hospitality of the Harvard-Smithsonian Center for
Astrophysics, where most of this work has been carried out.
This research has made use of Keck telescope time granted through NOAO
(program A285Hr).




\end{document}